\documentstyle[aps]{revtex}
\input epsf
\def\jepsfbox#1{\typeout{#1} \epsfbox{#1}}
\def\plotonesc#1#2{\begin{center} \leavevmode
\epsfxsize=#2\columnwidth \jepsfbox{#1} \end{center}}

\def\plottwo#1#2{\centering \leavevmode
\epsfxsize=.45\columnwidth \jepsfbox{#1} \hfil
\epsfxsize=.45\columnwidth \jepsfbox{#2}}
\def\jcite#1#2{#1 \cite{#2}}
\def\etal{{\it et al.\ }}

\def\cf{{\it c.f.~}}

\newcommand{\be}{\begin{equation}}
\newcommand{\ee}{\end{equation}}
\newcommand{\ba}{\begin{eqnarray}}
\newcommand{\ea}{\end{eqnarray}}
\def\figref#1{Fig.~\ref{fig:#1}}
\def\eqref#1{Eq.~\ref{eq:#1}}
\begin{document}
%
%
\title{Interfering with Interference: a Pancharatnam Phase Polarimeter}
\author{Jeremy S. Heyl}
\address{Theoretical
Astrophysics 130-33, California Institute of Technology,
Pasadena, California 91125}
\maketitle
\begin{abstract}
A simple variation of the traditional Young's double slit experiment
can demonstrate several subtleties of interference with polarized
light, including Berry and Pancharatnam's phase.
Since the position of the fringes depends on the polarization state of
the light at the input, the apparatus can also be used to measure the
light's polarization without a quarter-wave plate or an accurate
measurement of the light's intensity.  In principle this technique can
be used for any wavelength of photon as long as one can effectively
polarize the incoming radiation.
\end{abstract}

\section{Introduction}

\jcite{Pancharatnam}{Panc56} explored how the phase of polarized light
changes as the light passes through cycle of polarizations.  He found
that the phase increases by $-\Omega/2$, where $\Omega$ is the
solid angle that the geodesic path of polarizations subtends on the
Poincar\'e sphere.  If the path does not consist of great circles, an
additional dynamical phase will develop.  \jcite{Berry}{Berr84}
developed corresponding theory for general quantum systems and
re-derived Pancharatnam's result \cite{Berr87}.

A series of experiments have been performed which demonstrate the
Pancharatnam phase 
\cite{Bhan88a,Bhan88b,Simo88,Bhan90,Bhan92a,Bhan92b,Bhan93,Schm93} 
and the related geometric phase \cite{Chia86,Tomi86,Chia88}.
De Vito and Levrero \cite{Devi94} have criticized many of these
experiments because they use retarders which introduce a dynamical
component to the phase.  Berry \& Klein \cite{Berr96} and Hariharan
\etal \cite{Hari97,Hari99} have performed a series of
experiments using only polarizers and beam splitters to introduce and
measure the geometric phase.

In this paper, I describe a simple experiment using polarizers and a
double slit to demonstrate the Pancharatnam phase and use this phase
to determine the polarization state of the incoming light.  If the
incoming light is linearly polarized and the polarizers are also
linear, the resulting phase is limited to 0 or $\pi$.  However, 
elliptically polarized light will result in intermediate values of the
phase.  This experiment uses a double slit rather than a half silvered 
mirror to split the beam, eliminating a possible source of dynamic phase; 
furthermore, with fewer and less complex optical elements this experiment 
may be performed over a wide range of photon energies.  Furthermore,
since the optics are simple, one can analyze the experiment in terms
of Maxwell's equations, illustrating the connection between these
equations and the Pancharatnam phase.

\section{Experimental Apparatus}

\begin{figure}
\plotonesc{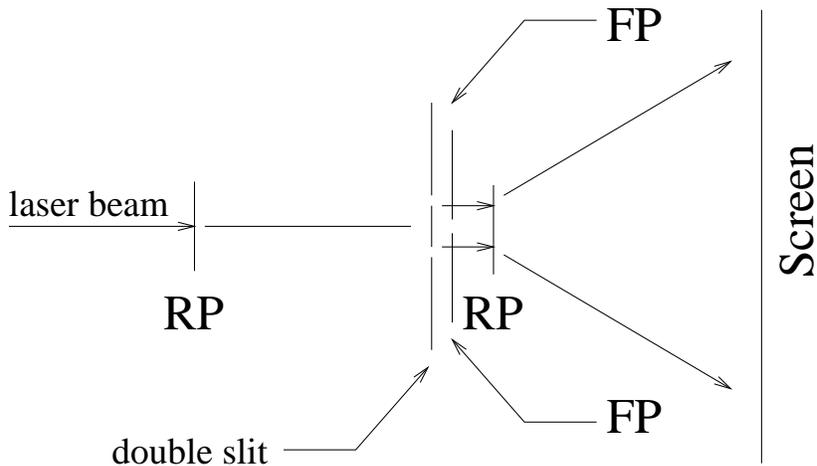}{0.6}
\caption{Optical system of the Pancharatnam phase polarimeter}
\label{fig:setup}
\end{figure}

\figref{setup} illustrates the experiment setup (\cf
\cite{Fort70,Hunt70,Pesc72}).  RP and FP denote rotatable 
and fixed linear polarizers respectively.  Each fixed linear
polarizer covers one of the two slits and is oriented perpendicular
to the other.  Each rotatable polarizer is oriented at forty-five degrees
to the polarizers to maximize throughput.  Incidentally, if the final 
polarizer is removed, the interference pattern
disappears according to the Frensel-Arago law \cite{Fort70}.

The arrangement is similar to that
used by Schmitzer, Klein \& Dultz \cite{Schm93}.  Instead of using a 
Babinet-Soleil compensator to vary the geometric phase. The phase depends 
on the input and output polarizations of interferometer.
If the two polarizers are aligned, the Pancharatnam phase between the two 
paths vanishes; and if they are orthogonal, the phase is $\pi$.

The setup is identical to the standard physics demonstration of
Young's double-slit experiment with the exception that each slit is
covered with a polarizing filter; consequently, both the construction
and analysis of the experiment are amenable as a demonstration or
student laboratory experiment.  Furthermore, the light follows the
same spatial trajectory, independent of the position of the polarizers
and the geometric phase observed.

\section{The Poincar\'e Sphere}

\begin{figure}
\plottwo{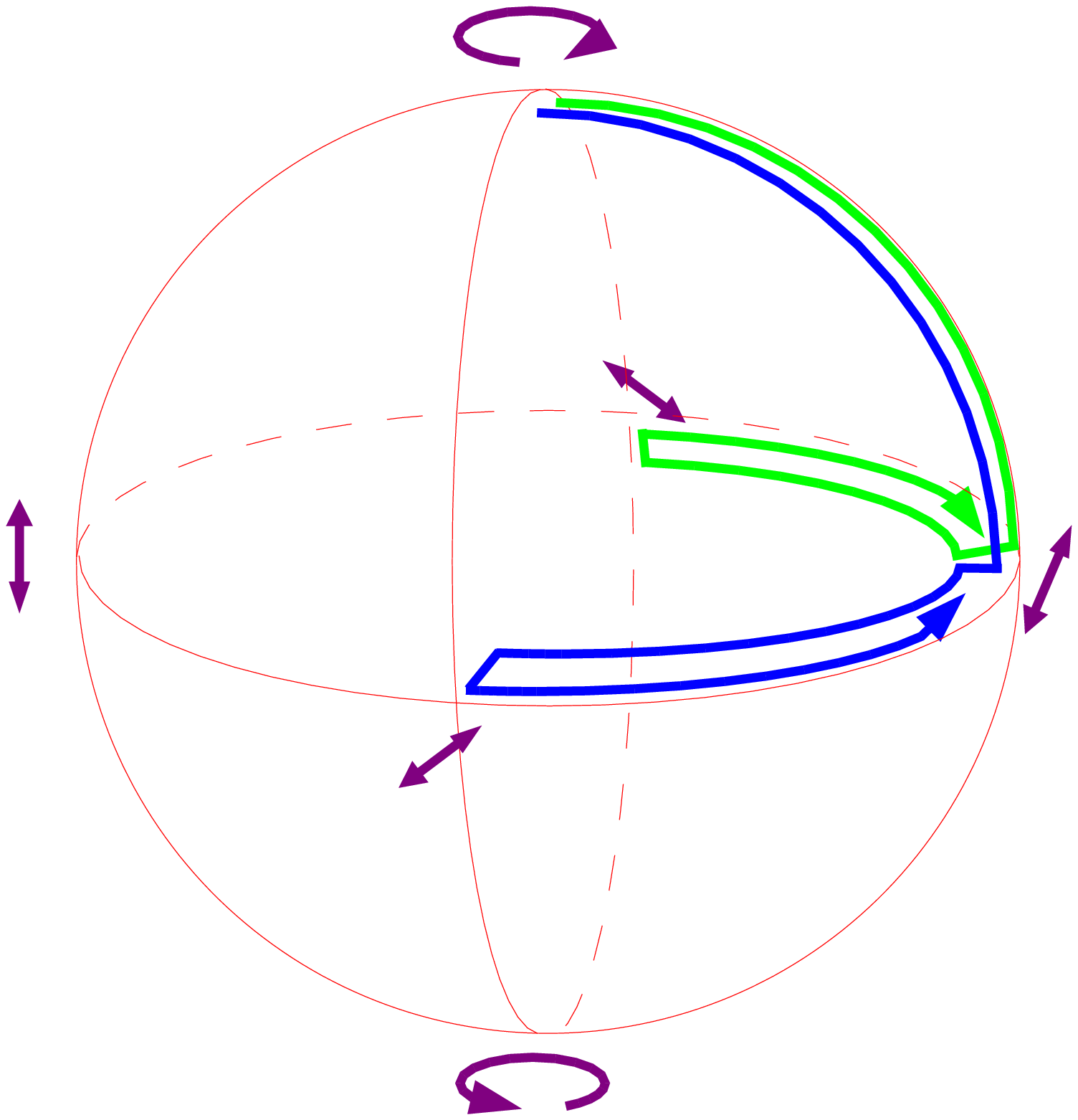}{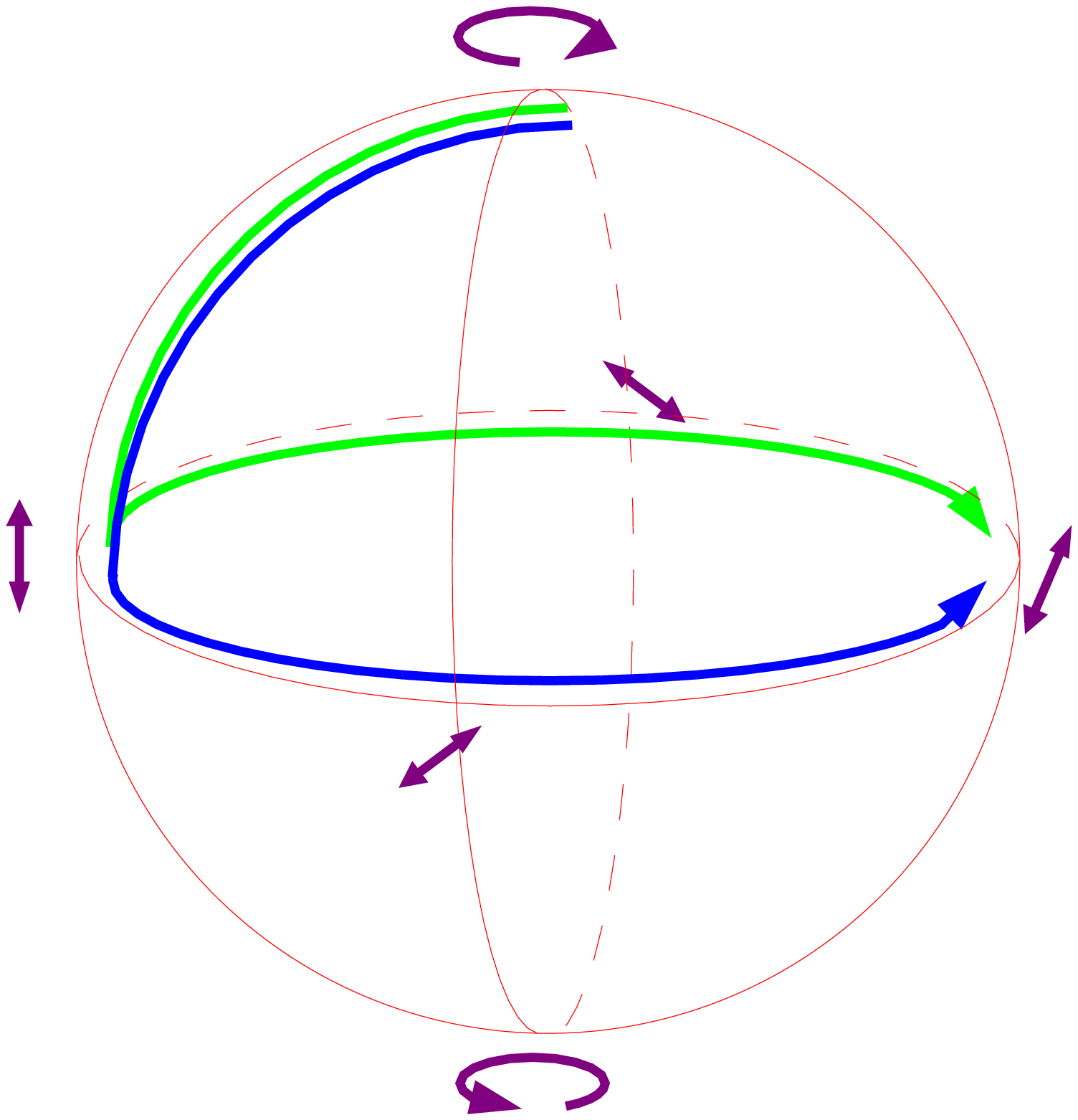}
\caption{The path the polarization follows on the Poincar\'e sphere for the 
input and output polarizers aligned in parallel (left) and
orthogonally (right).}
\label{fig:sphere1}
\end{figure}

Understanding how the apparatus works for a general polarization is
most simply achieved by tracing the polarization of the light through
the system along the Poincar\'e sphere.  The initial polarization from
the laser is unknown but it is depicted in the figures as left
circular.  The left panel of \figref{sphere1} depicts the
configuration for zero geometric phase.  As the laser passes through
the first polarizer, its polarization is projected onto the horizontal
direction.  After passing through the two slits, the polarization is
projected onto two orthogonal polarizations oriented at forty-five
degrees to the horizontal.  Finally, the last polarizer projects the
polarization back onto the horizontal.  One can form a closed loop by
following the path along one leg and returning along the other leg
(\cf\ \cite{Schm93}).  This closed loop does not enclose any solid
angle on the sphere.

The right panel of \figref{sphere1} illustrates the path of the
polarization when the two rotatable polarizers are orthogonal.  The
polarization is now projected onto the vertical first, followed by the
two diagonal polarizations and the final horizontal projection.
Constructing the closed loop as described earlier yields an area of
$2\pi$ and a Pancharatnam phase of $-\pi$ between the two slits.

Since a constant phase difference of $\pi$ is equivalent to $-\pi$,
this implementation hides the fact that the area on the sphere is
oriented and consequently the geometric phase may be positive or
negative.  A third configuration when the input polarizer is followed
by a quarter-wave plate yielding right-hand circular polarized input to
the interferometer illustrates this point.  The direction that the
loop is traversed determines which slit is designated by $\phi_1$
such that $\phi_1-\phi_2=-\Omega/2$.  The phase difference is given by
$\pi/2$ if the final polarizer lies clockwise relative to polarizer
behind the first slit, and $-\pi/2$ if the final polarizer lies
counterclockwise.  The converse result holds for left-hand polarized
light.

\begin{figure}
\plotonesc{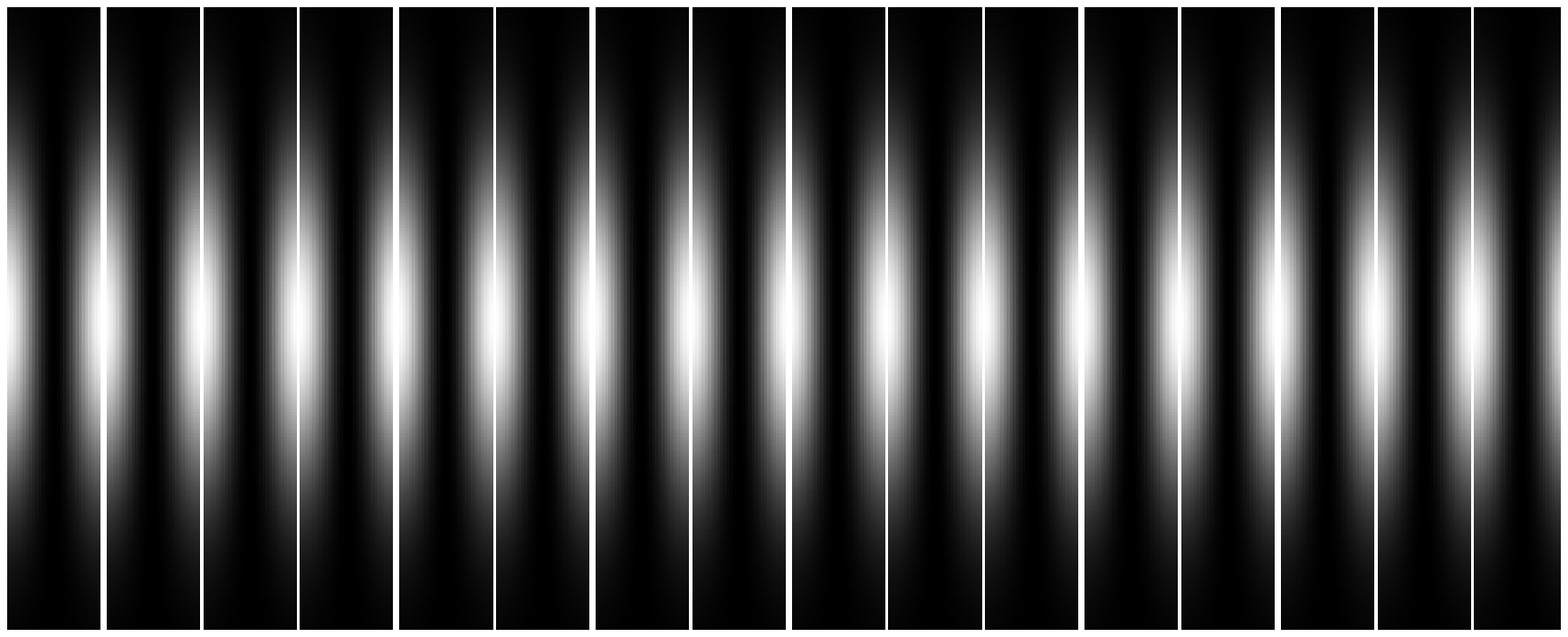}{0.5}
\plotonesc{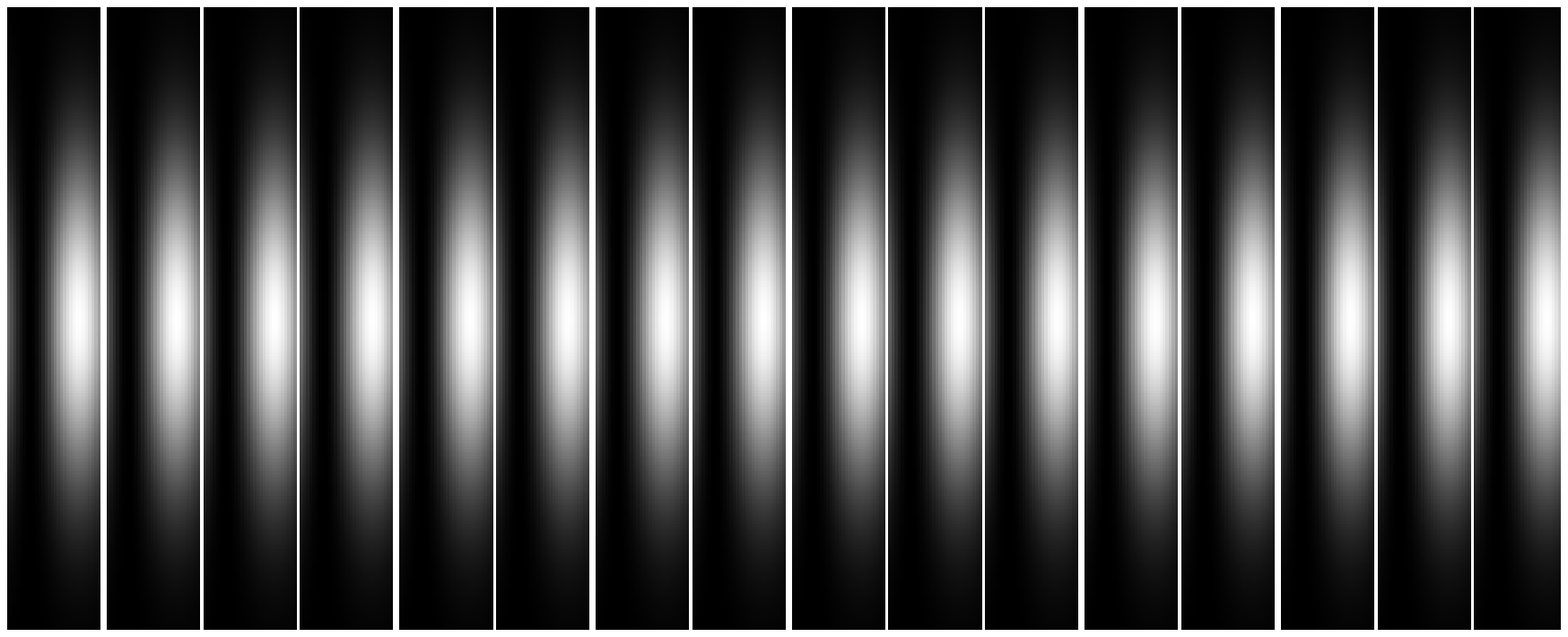}{0.5}
\plotonesc{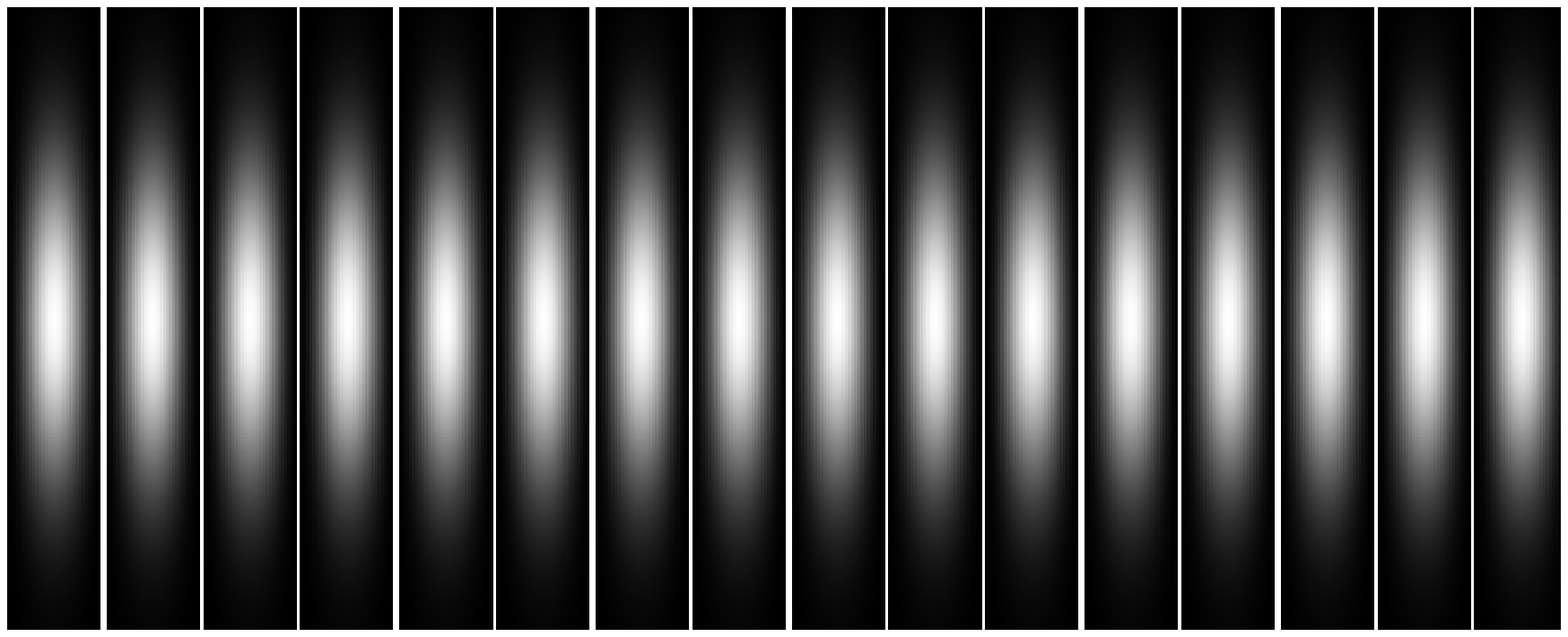}{0.5}
\caption{Schematic of the interference patterns for
$\phi_1-\phi_2=0,\pi/2,\pi$ for quasi-monochromatic light.}
\label{fig:pattern}
\end{figure}

If the axis of polarizer behind left-hand slit (as one looks toward
the screen) lies clockwise of that of the final polarizer, one
obtains the result that the fringes will shift to the left (relative
to their position for the input and output polarizations being
identical) if the light is left elliptically polarized and to the right
if it is right elliptically polarized.  This configuration
automatically includes the minus sign present in Pancharatnam's
definition of the geometric phase.  The upper and lower panels of 
\figref{pattern} show the
fringe pattern produced by the apparatus in the configuration
described above for the input and output polarizations being linear
and identical and linear and orthogonal.  The middle panel shows the
fringes for left-hand circularly polarized input light at the input 
with the input polarizer removed.

\section{Measuring the Input Polarization}

\begin{figure}
\plottwo{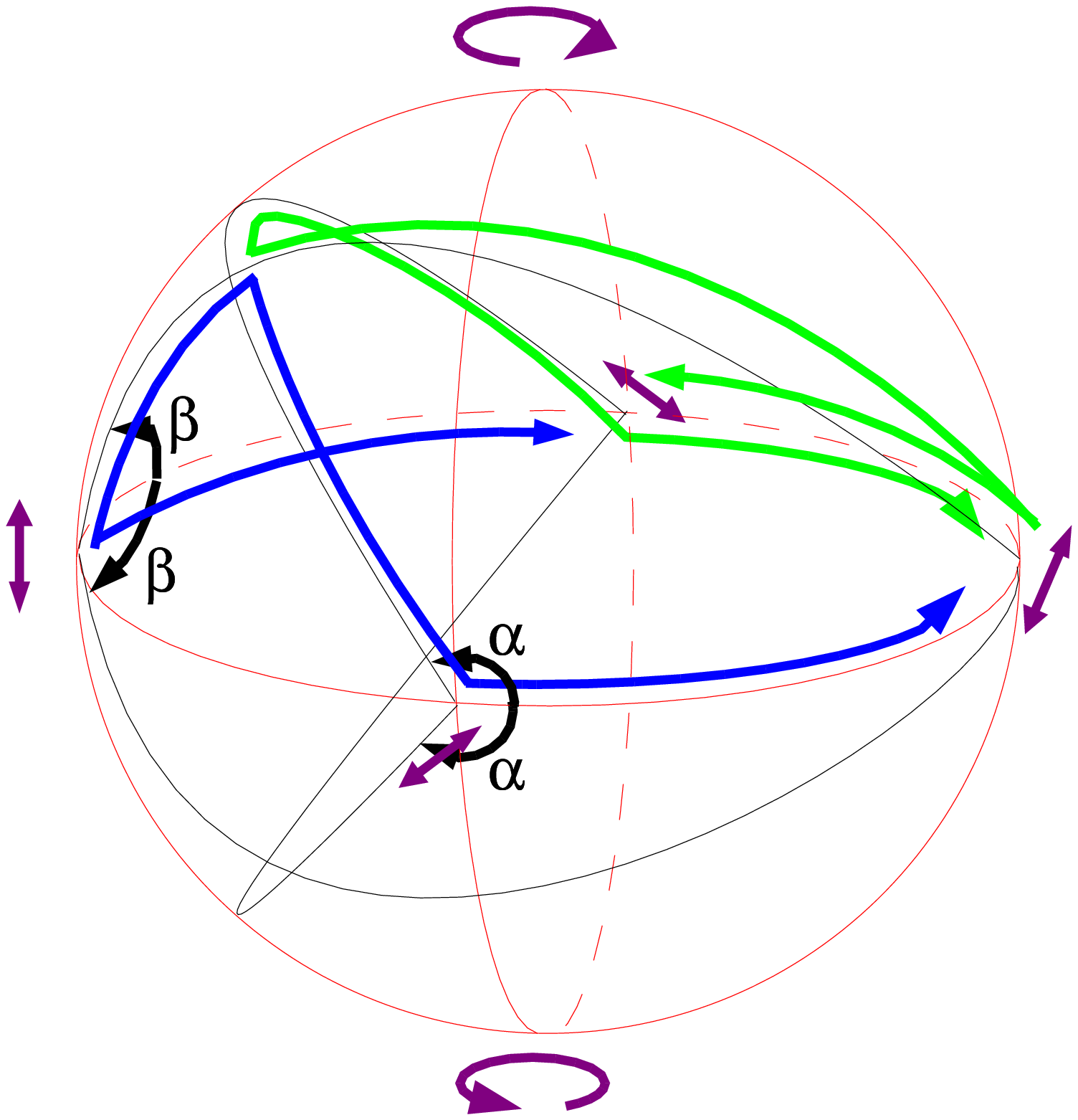}{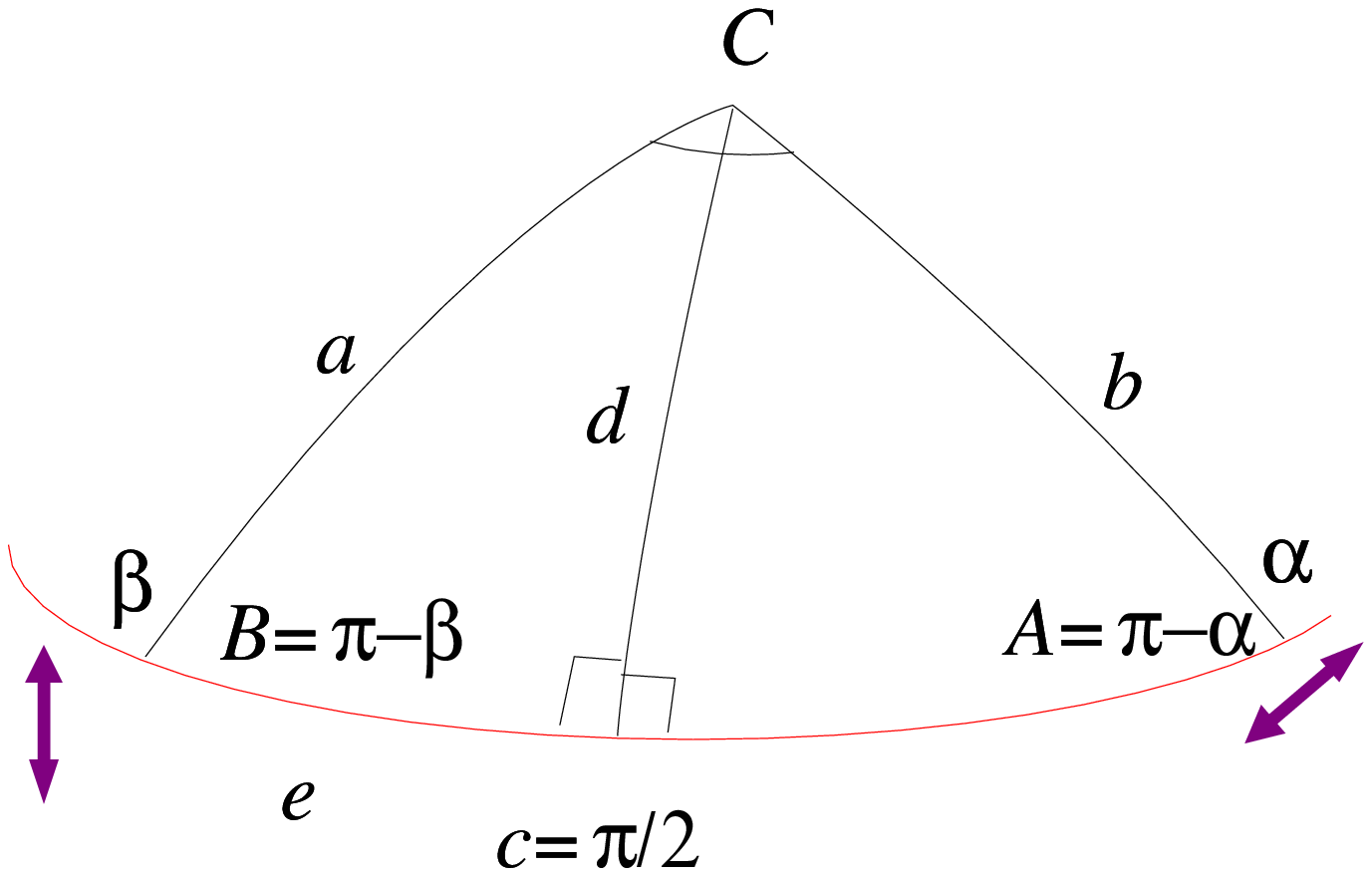}
\caption{Measuring and calculating the input polarization}
\label{fig:sphere4}
\end{figure}

If the input polarizer is removed, the apparatus can be used to
measure the polarization of the light source.
The procedure requires four measurements of the fringe positions: two
for calibration and two to determine the geometric phases.
\begin{enumerate}
\item
Locate the positions of the fringes with the 
input and output polarizers parallel and midway between the polarizers at 
the slits.
\item
Remove the input polarizer and compare the position of the fringes
relative to the two previous measurements.  Let $\alpha/(2\pi)$ be the
ratio of the offset of the fringes in step (2) relative to step (1) to
the distance between the fringes in step (1).  Also, note the direction that
fringes shift -- left for left-circular polarization and right for 
right-circular polarization. 
\item 
Rotate either all of the polarizers by forty-five degrees or rotate
the light source by forty-five degrees and repeat steps (1) and (2)
and denote the resulting ratio by $\beta/(2\pi)$.
\end{enumerate}
The left panel of \figref{sphere4} depicts how the polarization
evolves on the Poincar\'e sphere for the two configurations.  It is
straightforward to calculate the input polarization using spherical
trigonometry by following the right panel of \figref{sphere4}.  $\sin
d$ gives the fraction of circular polarization and $e$ is the angle
between the long axis of the polarization ellipse and the vertical axis.  
Napier's analogies yield
\be
a = - \left \{ \tan^{-1} \left [ \frac{\sin \frac{1}{2}
(\alpha-\beta)}{\sin\frac{1}{2} (\alpha+\beta)} \right ]
+ \tan^{-1} \left [ \frac{\cos \frac{1}{2}
(\alpha-\beta)}{\cos\frac{1}{2} (\alpha+\beta)} \right ] \right \}
\ee
and the law of sines gives
\be
\sin d = \sin a \sin \beta.
\ee
Combining these two results yields
\be
\tan \frac{e}{2} = \tan \frac{1}{2} (a-d) \frac{\sin\left
(\frac{\pi}{4} + \frac{\beta}{2}\right)}{\sin\left
(\frac{\pi}{4} - \frac{\beta}{2}\right)}
\ee

If the input polarization is linear, one finds that measurements of
the fringes can only locate the polarization vector to within
forty-five degrees.  However, the contrast of the interference pattern 
constrains the linear polarization further, as well as 
the fractional polarization of the input light.   

\section{Conclusions}

A variation of Young's double-slit experiment provides a excellent and
simple demonstration of the Pancharatnam phase for polarized
light.  Furthermore, the observed phase difference between the two
slits is simply related to the polarization of the incoming light.
The phase difference determines upon which great circle of the
Poincar\'e sphere the polarization lies, and by performing the
measurement after rotating the apparatus the input polarization can be
determined precisely.

The distinct advantage of the experiment is the simplicity of the
optics. Only linear polarizers and a double slit are required.  Since the 
Pancharatnam phase is achromatic, the procedure may be performed for the 
wide range of photon energies where suitable materials are available.  A 
companion paper discusses the implementation of the experiment in X-rays and
possible applications.

\acknowledgements 
I would like to thank Jackie
Hewitt, Lior Burko and Eugene Chiang for useful discussions and to
acknowledge a Lee A. DuBridge Postdoctoral Scholarship.

\bibliography{berry,ns,qed,leads,physics,mine} 
\bibliographystyle{prsty}

\end{document}